\documentclass{elsart}
\usepackage{graphics,graphicx,psfig,harvard,amssymb}
\journal{New Astronomy}
\input psfig.sty
\def\astrobj#1{#1} 

\begin{document}
\begin{frontmatter}

\title{New intrinsic-colour calibration for $uvby$--$\beta$ photometry}

\author[istanbul]{Y\"uksel Karata\c{s}\thanksref{1}},
\author[Ensenada]{William.J.~Schuster}
\thanks[1]{E-mail: karatas@istanbul.edu.tr}

\address[istanbul]{{\.I}stanbul University, Science Faculty, Department 
of Astronomy and Space Sciences, 34119 University, {\.I}stanbul, Turkey}

\address[Ensenada]{Observatorio Astron\'omico Nacional, Instituto de Astronom\'{\i}a,
Universidad Nacional Autonoma de M\'exico,\\
Apartado Postal 877, C.P. 22800, Ensenada, B.C., M\'exico}

\begin{abstract}
A new intrinsic-colour calibration ($(b$--$y)_{o}$--$\beta$) is presented
for the $uvby$--$\beta$ photometric system, making use of re-calibrated
Hipparcos parallaxes and published reddening maps.  This new calibration for
$(b$--$y)_{o}$--$\beta$, our Equation~(1), has been based upon stars with
$d_{Hip} < 70$ pc in the photometric catalogues of Schuster et al.~(1988,
1993, 2006), provides a small dispersion, $\pm0.009$, and has a positive
``standard''  $+2.239\Delta\beta$ coefficient, which is not too different
from the coefficients of Crawford (+1.11; 1975a) and of Olsen (+1.34; 1988).
For 61 stars with spectra from CASPEC, UVES/VLT, and FIES/NOT databases,
without detectable Na I lines, the average reddening value $\langle E(b-y) \rangle
= -0.001\pm0.002$ shows that any zero-point correction to our intrinsic-colour
equation must be minuscule.
\end{abstract}

\end{frontmatter}

\section{Introduction}

Intrinsic-colour, metallicity and absolute magnitude calibrations from
$uvby$--$\beta$ photometry for F-, G- and early K-type dwarf and turn-off stars
are of vital importance for studying several very important astrophysical
problems, such as age-metallicity relations, metallicity
gradients, and interstellar reddening in the Galaxy.  The $uvby$--$\beta$
photometric system was specifically devised by \citeasnoun{Stromgren} for
studying B, A, and F stars.  A recent evaluation of the importance and
capacity of the $uvby$--$\beta$ photometric system has been presented by
\citeasnoun{Bessell}.  For this photometry, several photometric indices, or colours,
are defined:  $(b$--$y)$, which measures the continuum slope and is sensitive to
stellar temperatures for B, A, F, and G stars; $m_{1}=(v$--$b)-(b$--$y)$, a colour
difference designed to measure the blanketing due to metal lines near
4100$\AA$ (this index is also referred to as a metal or metallicity index);
$c_{1}=(u$--$v)-(v$--$b)$, a colour difference designed to measure the strength of
the Balmer discontinuity; and $\beta=\beta_{w}-\beta_{n}$, an
intermediate-narrow index measuring the strength of the hydrogen $\beta$ line,
which is also sensitive to stellar temperatures for B, A, and F stars, and
which is free of the effects of interstellar extinction and reddening.

Since the $\beta$ index is independent of interstellar reddening, $(b$--$y)$ is
not, and both $\beta$ and $(b$--$y)$ measure stellar temperatures for B, A, F,
and G stars, this enables us to obtain an intrinsic-colour calibration,
$(b$--$y)_{o}$--$\beta$, which is quite important for de-reddening the
photometric indices $(b$--$y)$, $m_{1}$, and $c_{1}$ used for deriving the 
astrophysical parameters $T_{eff}$, $M_{v}$, and $[Fe/H]$ from $uvby$--$\beta$
photometry. For dwarf and turn-off stars, some of the first calibrations for
intrinsic-colour for $uvby$--$\beta$ photometry have been given by
\citeasnoun{Crawford75a} (hereafter C75), and \citeasnoun{Olsen88} (hereafter
O88).  C75 presents all three types of calibration (intrinsic-colour,
metallicity, and absolute magnitude) for Population I type stars in the solar
neighbourhood, with spectral types F2--G0 and luminosity classes III--V.  O88
gives only an intrinsic-colour ($(b$--$y)_{o}$--$\beta$) calibration good for
all F0--G2 stars of luminosity classes III--V, except perhaps the more metal-poor
Population II stars.  An intrinsic-colour calibration ($(b$--$y)_{o}$--$\beta$) was
also derived by \citeasnoun{SchusterNissen89} (hereafter SN89) in terms of the
standard (non-differential) indices $(b$--$y)$, $m_{1}$, $c_{1}$, and $\beta$.

Thanks to re-calibrated Hipparcos parallaxes (ESA 1997) by \citeasnoun{vanLeeuwen07},
the reddening maps of \citeasnoun{Schlegeletal} (hereafter SFD98), and the large
photometric data base of \citeasnoun{SchusterNissen88} (hereafter SN88),
\citeasnoun{Schusteretal93} (hereafter SPC93), and \citeasnoun{Schusteretal06}
(hereafter SMM06) a new relation for intrinsic-colour ($(b$--$y)_{o}$--$\beta$)
based on $uvby$--$\beta$ photometry for F-, G-, and early K-type dwarf and turn-off
stars has been obtained.  The calibration is updated and improved compared to those
which have been available in the literature and used in $uvby$--$\beta$ photometric
surveys to derive abundance distributions, effective temperatures, age-metallicity
relations, and stellar kinematics for the Galaxy (for example, see
\citeasnoun{NissenSchuster91},  SPC93, SMM06, and \citeasnoun{Nordstrometal}).
This new intrinsic-colour calibration also has the advantage that its zero-point is
tested here using stars observed spectroscopically with the CASPEC, UVES/VLT, and
FIES/NOT echelle spectrographs and shown to have no interstellar Na I lines.

This paper is organized as follows:   Section~2 describes the data catalogues,
parallaxes, and cleansing of binary stars.  In  Section~3 the de-reddening procedure
is presented in detail. In Section 4 the $(b$--$y)_{o}$--$\beta$ relation from the
Schuster et al. photometric catalogue is presented, some comparisons are made, a
discussion of these, and finally the conclusions in Section~5.

\section{Data, and Removal of binaries}

Our new $(b$--$y)_{o}$--$\beta$ calibration has been based mainly on the
$uvby$--$\beta$ catalogues of \citeasnoun{SchusterNissen88}\footnote{Based on
observations collected at the H.~L. Johnson 1.5m telescope at the
Observatorio Astron\'omico Nacional at San Pedro M\'artir, Baja California,
M\'exico, and at the Danish 1.5m and 0.5m telescopes at La Silla, Chile.}
(hereafter SN88), and Schuster et al.\footnote{Based on observations
collected at the H.L. Johnson 1.5m telescope at the Observatorio
Astron\'omico Nacional at San Pedro M\'artir, Baja California, M\'exico}
(SPC93; SMM06).  First, a catalogue which includes 1475 dwarf and turn-off
stars from the SN88, SPC93 and SMM06 catalogues has been created, and
this $uvby$--$\beta$ catalogue will be referred to as the ``Schuster'' 
catalogue.  Photometric ranges for this catalogue are as follows:
$0.159 \leq (b$--$y) \leq 0.728$, $0.033 \leq m_{1} \leq 0.757$, 
$0.088 \leq c_{1} \leq 0.842$, $2.486 \leq \beta \leq 2.792$.
Particular importance has been given to excluding possible binary,
variable, and flare stars from the Schuster catalogue.  Stars which have
comments and notes, such as (N,D), (N, D*), (S, D), (S, D*), (N/S, D),
(N/S, D*), (S, HM), R, and R? in the SN88 catalogue have been excluded.
In general, these are stars observed with a second fainter star within or
slightly outside the entrance diaphragm of the photometer (``D''), stars
with $\beta$ values taken from an outside source (``HM''), or stars near or
outside the color limits of our photometric transformations (``R'' or ``R?'').
Similarly for the catalogue of SPC93, and in addition stars labelled with
``++'' have been omitted; these are the redder subgiant/giant stars whose
$uvby$--$\beta$ photometry probably contains small systematic
transformation errors.  Stars with ``+'' from this same SPC93 catalogue
are not as red, $(b$--$y) < 0.50$, and have been retained.  Also, for
the catalogue of SMM06 stars with similar binary indications in the notes,
as well as ``++'', have been excluded, and in addition those with the
note ``fainter star in diaphragm'', but stars with ``fainter star (just)
outside diaphragm'' have been retained.

Since binaries and anomalous stars have an impact not only on the
intrinsic-colour calibration, but also on the metal-abundance, and
absolute-magnitude calibrations, the Schuster catalogue has been cleansed
of these type of stars.  SMM06 identified binaries by using various
catalogues, including primarily those of \citeasnoun{Carneyetal94}, 
\citeasnoun{Carney03}, and \citeasnoun{DommangetNys}.  235 stars which are
doubled-lined spectroscopic binaries (SB2), other types of binaries, or
photometrically variable stars have been removed from this Schuster catalogue.

Re-calibrated Hipparcos parallax data by \citeasnoun{vanLeeuwen07} and
$E(B$--$V)$ reddening values from SFD98 have been collected for this
$uvby$--$\beta$ Schuster catalogue.  For this intrinsic-colour calibration, 
parallaxes ($\pi$), their associated uncertainties ($\sigma_{\pi}$), and
Galactic coordinates ({\it l}, {\it b}) of stars have been taken from
\citeasnoun{vanLeeuwen07}.

\section{Interstellar Reddening}

Understanding the local interstellar reddening is of vital importance
in the derivation of the $(b$--$y)_{o}$--$\beta$ calibration, and likewise
for stellar metallicities, distances and ages.  The region which lies at
50--100~pc from the Sun is specially interesting.  The nearest interstellar
dust patches are at about 70~pc in some directions; this region devoid of dust
is commonly identified with the Local Hot Bubble.  There are several works
about reddening within the limits of this Local Bubble: very weak interstellar
polarization caused by magnetically aligned dust grains has perhaps been
observed within $\approx $35 pc by \citeasnoun{Tinbergen}, but \citeasnoun{Leroy}
finds almost no polarization up to 50~pc.  \citeasnoun{Leroy} also showed that
significant dust clouds appear at 70--80~pc, slightly beyond the Local Bubble
boundary defined with the help of X-ray measurements.  \citeasnoun{Holmbergetal}
consider that real reddening within 40~pc is negligible.  The older
intrinsic-colour calibration of SN89 was derived with a distance limit of about
80~pc, and an over-correction for interstellar reddening of about $+0.005$ was
suggested by \citeasnoun{Nissen94} using stars without detectable interstellar
Na I lines.  Additionally,  \citeasnoun{Vergely}, \citeasnoun{Sfeiretal},
\citeasnoun{Breitschwerdtetal}, \citeasnoun{Lallementetal},
\citeasnoun{LuckHeiter} state that the extinction within 65--75~pc of the Sun
is essentially nil.  From all these works concerning the Local Bubble, the
reddening values within 70~pc are seen to be insignificant.

To derive the intrinsic-colour ($(b$--$y)_{o}$--$\beta$) calibration, as a
first step, only stars with relative-parallax errors, $\sigma_{\pi}/\pi \leq 0.10$,
are considered, corresponding to the near-solar vicinity,
where interstellar reddenings are small or negligible.
To check the distribution of reddening in the solar neighbourhood,
$E(B$--$V)(l, b)_\infty$ values for individual stars in the Schuster and
Olsen catalogues are taken from the reddening maps of SFD98 via the web
page of NED (NASA Extragalactic Database).  The reddening $E(B$--$V)$ for a
given star is reduced compared to the total reddening $E(B$--$V)(l, b)_\infty$
by a factor $\lbrace1-exp[-d \sin |b|/H]\rbrace$, where {$b$} and {$d$} are
the Galactic latitude and distance, respectively, assuming that the dust layer
has a scale height $H = 125$ pc \cite{Bonifacioetal}.  However,
\citeasnoun{ArceGoodman} caution that the SFD98 reddening maps overestimate the
reddening values when the colour excess $E(B$--$V)$ is more than about $0.15$,
or even as low as $0.10$ \cite{Schusteretal04}.  Hence, according to
\citeasnoun{Schusteretal04}, a slight revision of the SFD98 reddening estimates
has been adopted via an equation, $E(B$--$V)_{\rm A} = 0.10 + 0.65(E(B$--$V)-0.10)$
when $E(B$--$V) > 0.10$, otherwise $E(B$--$V)_{\rm A} = E(B$--$V)$,
where $E(B$--$V)_{\rm A}$ indicates the adopted reddening estimate.

Graphs of $E(B$--$V)$ versus Hipparcos distance ($d_{Hip}$) and versus Galactic 
latitude ($b$) for 551 stars with $\sigma_{\pi}/\pi \leq 0.10$ from the 
Schuster catalogue are plotted in Figs.~1(a) and 1(b), respectively, for the
Galactic latitude range of $0^{o} \leq |b| \leq 90^{o}$.  Figs.~1(a)--(b) allow
us to appreciate the reddening in the solar vicinity for the derivation of the
$(b$--$y)_{o}$--$\beta$ calibration.  It can be seen from Fig.~1(a) that the
majority of stars have reddening values less than $E(B$--$V)\approx 0.02$
(the horizontal line), and most reddenings are small up to
distances of $d_{Hip} \approx 70$ pc.  $E(B$--$V)$ = 0.02 corresponds approximately to
$E(b$--$y) = 0.012$ from the relation $E(B$--$V) = 1.35E(b$--$y)$ from
\citeasnoun{Crawford75b}.  Although there are a few stars with reddenings up to
$E(B$--$V) \approx 0.10$ in Figs.~1(a)--(b), the reddening values of
most stars are less than $0.02$ for $0^{0} \leq |b| \leq 90^{0}$.
In Figs.~1(a)--(b) strict limits are put on the reddening in the solar
vicinity for the derivation of our intrinsic-colour calibration.  
\citeasnoun{MV}(their figure~3) show that reddening values for three low-latitude directions are
zero for distances less than about 100 pc.  SN89 consider that reddening is
negligible within a distance limit of 100 pc following the works of C75 and
\citeasnoun{Crawford79}; their actual distance limit may be more like 80~pc from a
more exact photometric distance scale.  Olsen used the criteria $d < 70$ pc
and $E(b$--$y) < 0.03$ for his intrinsic-colour calibration.

Metal-poor stars ($[Fe/H] < -1.00$ dex) are of vital importance for the
intrinsic-colour calibration from the Schuster catalogue, since this
calibration is intended for a full range of stellar population types, from thin
disk to extreme halo. Having taken into consideration the distance limit for
the reddening of Local Bubble, reddening is considered to be negligible for stars
with $d_{Hip} < 70$~pc from the distributions displayed in Figs.~1(a)--(b), and
according to the references mentioned above.  According to these publications
concerning the Local Bubble, this $d_{Hip} < 70$ pc criterion, which is slightly
beyond the Local Bubble boundary, can be justified as an upper limit.  Also, a
smaller distance limit would leave too few metal-poor stars for an adequate
intrinsic-colour calibration.

In summary, the strict criteria from Figs.~1(a)--(b), plus reddening studies 
concerning the Local Bubble, allow us to derive an intrinsic-colour
calibration using stars from the Schuster catalogue out to distances of
70~pc, with $E(B$--$V) < 0.02$.

\section{The Intrinsic-colour Calibration from the Schuster Catalogue}

The $(b$--$y)_{o}$--$\beta$ calibration has been carried out utilizing 
a mathematical package called ``Minitab'' which allows the regression
of a dependent variable against several independent variables.  From the
Schuster catalogue, 405 stars with $E(B$--$V) < 0.02$ and 
$d_{Hip} < 70$ pc, according to panels~(a) and (b) of Fig.~1,
have been used for the derivation of the present intrinsic-colour
calibration.  (These calibration stars are assumed to be unreddened, and
so $m_{1} = m_{o}$ and $c_{1} = c_{o}$.)  For these 405 stars, sixteen
terms from simple ones to higher-order cross terms have been tested, as
follows:  $m_{o}$, $m_{o}^2$, $c_{o}$, $c_{o}^2$, $\Delta\beta$,
$\Delta\beta^2$, $m_{o}\Delta\beta$, $m_{o}c_{o}$, $m_{o}c_{o}^2$,
$m_{o}^2c_{o}^2$, $c_{o}\Delta\beta$, $m_{o}\Delta\beta^2$,
$m_{o}^2\Delta\beta$, $c_{o}^2\Delta\beta$, $c_{o}\Delta\beta^2$, and
$m_{o}c_{o}\Delta\beta$, where $\Delta\beta$ = $2.720-\beta$.  The t-ratios,
the ratios between a coefficient and its estimated error, were used to
eliminate non-significant terms. The solutions were iterated; at each step
the term with the smallest t-ratio was omitted until all terms were
significant.  During the Minitab regression analyses, 13 stars have been
removed, those with residuals greater than $\pm0.025$, since they are
probably slightly reddened, for the positive cases, or somewhat anomalous,
for the negative.  The remaining data have 375 degrees of freedom and so all
coefficients are non-zero at a significance level greater than 0.982, since
the solution has been iterated until all terms have t-ratios with absolute
values greater than 2.59.  These terms were removed while doing the
subsequent iterations:  $m_{o}\Delta\beta$, $c_{o}^2$, $m_{o}^2\Delta\beta$,
$m_{o}$, $m_{o}^2$, and  $m_{o}^2c_{o}^2$.

The final solution, with a dispersion of $\pm0.0088$, is given as follows:

\begin{eqnarray}
(b-y)_{o}=+0.492(\pm0.07)-0.976(\pm0.14)c_{o}\nonumber\\ 
+2.239(\pm0.77)\Delta\beta-8.77(\pm3.01)\Delta\beta^2\nonumber\\ 
+6.26(\pm0.39)m_{o}c_{o}-16.51(\pm2.79)c_{o}\Delta\beta\nonumber\\ 
-4.720(\pm0.80)m_{o}c_{o}^2+53.24(\pm10.61)c_{o}\Delta\beta^2\nonumber\\ 
+9.39(\pm1.65)m_{o}\Delta\beta^2+27.526(\pm3.03)c_{o}^2\Delta\beta\nonumber\\  
-26.757(\pm2.41)m_{o}c_{o}\Delta\beta  
\end{eqnarray}

\noindent Equation (1) is valid for the ranges, $+0.290 \leq (b$--$y) \leq +0.606$,
$+0.060 \leq m_{1} \leq +0.574$, $+0.126 \leq c_{1} \leq +0.504$, and
$2.488 \leq \beta \leq 2.668$, has a small dispersion, $\pm 0.0088$, and
contains a positive $\Delta\beta$ coefficient, $+2.239$, which
is not too different from the coefficients of Crawford (+1.11; 1975a) and
of Olsen (+1.34; 1988). In Table~1 average residuals for
some sub-groups of $m_{1}$, $c_{1}$, and $\beta$ are listed for Equation~(1);
average residuals, numbers, and standard deviations from Equation~(1) are
listed in Columns~2, 3, and 4, respectively.  Note that the ranges in the
average residuals for $m_{1}$, $c_{1}$, and $\beta$ for Equation~(1) are
$[-0.0013, +0.0007]$, $[-0.0014, +0.0015]$, and $[-0.0022, +0.0009]$, respectively.

The intrinsic-colour calibration of SN89 was derived from 267 stars and retains
nine terms, including an $m_{o}$ term, whereas the new calibration in Equation~(1),
392 stars and 11 terms, without $m_{o}$, but the $m_{o}$ dependence is available
in a more complicated form, in four cross terms: $m_{o}c_{o}$,
$m_{o}c_{o}^2$, $m_{o}\Delta\beta^2$, and $m_{o}c_{o}\Delta\beta$; during the
iterations of the calibration process, the simple $m_{o}$ term has been eliminated
according to the empirical criteria discussed above.  Here, sixteen terms have been
considered while producing the new intrinsic-colour calibration as compared to
only twelve terms for the one of SN89.  Moreover, the number of calibration stars
in Equation~(1) and the metal-poor-star content is superior to that of SN89, and
finally this new intrinsic-colour calibration extends to somewhat cooler stellar
temperatures as shown by the applicable ranges in $(b$--$y)$, $m_{o}$, and $\beta$
for these two calibrations; stellar spectra are becoming more complicated at cooler
temperatures.

$E(b$--$y)$ distributions for stars from the Schuster catalogue are displayed in
Figs.~2(a) and (b) as calculated from Equation~(1); the stars included fall in
the ranges of validity given above.  Panel~2(a) includes 1062 stars within the
above limits of $(b$--$y)$, $m_{1}$, $c_{1}$, and $\beta$ in the Schuster catalogue,
while 2(b) only those stars with $[Fe/H] < -1.0$.  (For program stars which may be
reddened, $m_{1}$ is substituted for $m_{o}$ and $c_{1}$ for $c_{o}$ in Equation~(1),
and then the solution is iterated to consistency in $(b$--$y)_{o}$, $m_{o}$, and
$c_{o}$; see SN89.)  The hatched areas of both panels in Fig.~2 fulfill the distance
limit $d_{Hip} < 70$ pc, which corresponds to the near-solar vicinity and negligible
reddenings, as argued above.  Qualitative agreements can be seen in Figs.~2(a) and
(b); all open histograms have positive tails extending to $E(b$--$y) \approx 0.07$,
showing slightly reddened stars within distributions which are mostly symmetric
about $E(b$--$y) = 0.00$.  As expected, the hatched histograms of Fig.~2 show
$E(b$--$y)$ distributions mostly symmetric about $E(b$--$y) = 0.00$, and no obvious
reddening tails; this is true for both the sample of 515 stars including all stars
in panel~(a), and for the 21 stars with lower metallicities of panel~(b).
In fact, the 21 hatched, low-metallicity stars all have $E(b$--$y)$ less than
$0.03$.

For the distance $d_{Hip} < 70$ pc and metallicities $[Fe/H] < -1.0$,
there are 21 stars in the Schuster catalogue; mean $\langle E(b$--$y) \rangle$ values
for these metal-poor stars are presented in Table~2.  $\langle E(b$--$y) \rangle$
values are $+0.004$ and $+0.010$ from Equation~(1) for $[Fe/H]_{spec} < -1.0$, and
$[Fe/H]_{spec} < -1.5$, respectively.

In Table~3, Column~3, $\langle E(b$--$y) \rangle$ values from Equation~(1) are compared
with those from SN89, or from O88, for different metallicity groups within the limits of
these calibrations.  Here, $\langle\Delta E(b$--$y)_{eq.(1)- SN89}\rangle =
E(b$--$y)_{eq.(1)}- E(b$--$y)_{SN89}$, and correspondingly for
$\langle\Delta E(b$--$y)_{eq.(1)- O88}\rangle$.  When our $E(b$--$y)$ values are compared
to those of SN89, the zero point in the Equation~(1) of SN89 has been taken both
with and without the small zero-point correction of $+0.005$ discussed by Nissen (1994)
(hereafter N94), based on the $E(b$--$y)$ values of 23 stars with undetectable Na~I
lines; the left half of Column~3, Table~3, gives the $\langle\Delta E(b$--$y)\rangle$
differences with this correction and the right half (in parentheses), without.  The mean
$\langle\Delta E(b$--$y)\rangle$ differences are quite small, as can be seen in the
left half of Column~3, and are $0.005$ smaller without this zero-point correction.
Note that the average $\Delta E(b$--$y)$ differences show no systematic trend with $[Fe/H]$.

For the comparisons of Table~3, $E(b$--$y)$ values are estimated for stars in the Schuster
catalogue considering the appropriate limits of the intrinsic-colour equations of SN89 and
O88. The equation of O88 is valid for the ranges of $\delta c_{1}$ = [$-0.02$, $+0.25$]
and $\delta m_{1}$ = [$-0.01$, $+0.135$], whereas the relation of SN89 is valid for
$(b$--$y)$ = [0.254, 0.550], $m_{1}$ =[0.033, 0.470], $c_{1}$= [0.116, 0.540], and
$\beta$ = [2.550, 2.681].  SN89 showed that the average difference of $E(b$--$y)$
between their calibration and O88 had a systematic trend with metallicity, and confirm
the need of a correction of $+0.015$ mag over $-2.5 < [Fe/H] < -1.5$ for the O88
calibration.  In Table 3, the trends of $\Delta E(b$--$y)$ between Equation~(1) and
O88 agree with those in table~4 of SN89.  In the lower part of Table~3, at high
metallicity, there is a non-negligible ($\approx -0.02$) difference between the $E(b$--$y)$
values of Equation~(1) and those from O88, and this varies by almost 0.02 mag in passing
from the metal-rich to the metal-poor regime; such a variation of about 0.02 mag is also
seen in the comparison of SN89 (table~4).  Both comparisons in the lower part of Table~3
have negative values, suggesting an overestimation of the reddening by O88, while the
comparison to SN89 suggests a small underestimation by their calibration.  

Our intrinsic-colour calibration, Equation~(1), is based in part on the 
COBE/ DIRBE, IRAS/ISSA full-sky dust maps of SFD98, by using the selection 
criteria of $E(B$--$V) < 0.02$, plus the distance criteria of $d < 70$ pc, 
derived by considering the reddening values within the Local Bubble,
while the corresponding calibration of SN89 depends mostly on a distance criterion,
$d \lesssim 80$ pc.  Both calibration procedures reject calibrating stars with residuals
greater than $\pm0.025$ during the iterations.  The $+0.005$ zero-point correction
of Nissen94 is based on the histogram of his figure~2, which shows an expected
rms scatter in $E(b$--$y)$ of $\pm 0.009$, the maximum bar centred at
$E(b$--$y) = 0.00$, but a slight asymmetry leading to a small positive
$\langle E(b$--$y) \rangle = +0.005$ for the 23 metal-poor stars observed with the
ESO 3.6m telescope and its CASPEC echelle spectrograph and showing no interstellar
Na I lines.

Such a procedure assumes that stars without interstellar Na lines are not affected
by interstellar extinction, i.e. $E(b$--$y) = 0.00$.  \citeasnoun{MunariZwitter}(their fig.~4)
have shown there is a good correlation between EW(Na I) and E(B-V) for single-lined systems.
The works of \citeasnoun{Hobbs}, \citeasnoun{Sembachetal}, \citeasnoun{SembachDanks}, 
and N94 have also shown correlations between the interstellar 
Na I gas and dust.  In fact, equation~(2) of \citeasnoun{Hobbs}, equation~(1) of  
N94, and figure~4 of  \citeasnoun{MunariZwitter}
all point to the probability that the interstellar dust abundance goes to zero 
together with the interstellar Na I gas abundance.  These works point out that cold gas and
dust probably occur together, and that the absence of cold interstellar gas implies the
absence of dust.  These results strengthen greatly our assumption that the absence of
interstellar Na~I lines proves the absence of interstellar dust, i.e. $E(b$--$y) = 0.00$.

To be able to check the zero-point of our new intrinsic-colour equation, the 23 metal-poor
CASPEC stars are still available, and also UVES/VLT halo and thick disk stars, as well as 
FIES/NOT halo stars, as provided by Prof. P.~E. Nissen.  The UVES/VLT spectra have been
taken from the ESO/ST-ECF Science Archive Facility, and the FIES/NOT halo stars were
observed at the Observatorio del Roque de los Muchachos by  Nissen and Schuster during two
observing runs in 2008.  A total of 61 different stars are available, all observed with
echelle spectrographs, and all selected to show no interstellar Na I lines in their
high-resolution spectra.

$E(b-y)$ values for the 23 CASPEC stars are presented in panel~(a) of Fig. 3; these are
calculated via Equation~(1) using each star's values of $(b$--$y)$, $m_{1}$, $c_{1}$, and
$\beta$.  Panel~(a) gives $\langle E(b-y) \rangle = +0.004\pm0.002$ (mean error), which is
very close to the value of $+0.005$, the zero-point correction of N94.  The comparisons of
Table~3, plus the results of \citeasnoun{Nissen94}, would suggest that the selection
criteria of these two intrinsic-colour calibrations (our Equation (1) and equation (1) of
SN89), especially the rejection of calibrating stars with residuals greater than $\pm0.025$,
will tend to lead consistently to intrinsic-colour equations which slightly over-correct
for the interstellar reddening.  Or, that the analysis of N94, and that of our Equation (1),
have perhaps been affected by small-number statistics; only about seven stars cause the
asymmetry of his figure~2.

In the new UVES/VLT and FIES/NOT databases, provided by Prof. P.~E. Nissen, 44 F- and G-type
stars without detectable interstellar Na~D lines are available with $(b$--$y)$, $m_{1}$,
$c_{1}$, and $\beta$ photometry.  These give the $E(b-y)$ distribution in panel~(b) of Fig.~3,
which shows an average reddening value of $\langle E(b-y) \rangle = -0.003\pm0.003$ (mean error)
from our Equation (1), and this value is insignificantly different from $E(b$--$y) = 0.00$.

Combining these three samples, excluding a few overlaps in the CASPEC, UVES/VLT, and FIES/NOT
databases, and including the thick-disk, spectroscopic standard stars HD 22879 and HD 76932,
also without interstellar Na I lines \cite{NissenSchuster09}, the resulting $E(b-y)$
distribution for 61 stars is presented in panel~(c) of Fig.~3.  The average reddening value
$\langle E(b-y) \rangle = -0.001\pm0.002$ (mean error) is obtained, which again is
insignificantly different from $E(b$--$y) = 0.00$.  These new F- and G-type metal-poor stars
without interstellar Na I lines suggest that any zero-point correction to our intrinsic-colour
equation must be very small.

Our new intrinsic-colour calibration has been applied to the main-sequence stars of the open
clusters \astrobj{NGC~2548} and \astrobj{M67}, which have CCD $uvby$--$\beta$ photometry
published by \citeasnoun{Balagueretal05} (hereafter B05) and \citeasnoun{Balagueretal07}
(hereafter B07).  For 124 main sequence stars of the M67 cluster within the applicable
limits of Equation~(1), our intrinsic-colour calibration gives the average reddening
$\langle E(b-y) \rangle = +0.020\pm0.004$.  B07 give $+0.030\pm0.030$ for \astrobj{M67},
derived using standard photometric relations to obtain the stellar parameters, as described
in  \citeasnoun{Jordietal}.  Also for \astrobj{M67}, \citeasnoun{Nissenetal87}(hereafter N87)
obtained $\langle E(b-y) \rangle = +0.023\pm0.004$ from $uvby$--$\beta$ photometry of
main-sequence stars, again showing very good agreement; they used several methods of
C75, \citeasnoun{Crawford78}, \citeasnoun{Crawford79} and of \citeasnoun{Hilditchetal},
depending on the spectral range of the main sequence stars.  For 21 main sequence stars of
the open cluster \astrobj{NGC~2548}, which fall within the applicable limits of our
Equation~(1), our calibration gives the average reddening
$\langle E(b-y) \rangle = +0.060\pm0.011$, while B05 obtain $+0.060\pm0.030$, again using
the method of standard photometric relations.  Our estimated average reddenings for these
two clusters are in good concordance with the ones found by B05 and B07, and by N87.

\section{CONCLUSIONS}

Our main conclusions are as follows:

\begin{enumerate} 

\item The intrinsic-colour calibration of our Equation~(1) has a small
dispersion, $\pm0.0088$, and its $+2.239\Delta\beta$ term is a standard
positive one like those of Crawford (+1.11; 1975a) and Olsen (+1.34; 1988).
Equation~(1) also has the advantage of being useful over a wider range of
stellar-population types, from metal-rich to quite metal-poor ($-2.72\leq[Fe/H]\leq+0.42$).

\item The $E(b-y)$ distribution of 23 CASPEC stars, as can be seen from panel~(a)
of Fig. 3, shows $E(b-y) = +0.004\pm0.002$, which is almost the same as the
$+0.005$ zero-point correction of N94.  As discussed by Nissen,
this suggests that the zero point of our intrinsic-colour equation should be
increased by $+0.004$.

\item However, for 44 stars of the UVES/VLT and FIES/NOT databases without detectable
interstellar Na I lines, the average reddening value $\langle E(b-y) \rangle = -0.003
\pm0.003$, with a sign opposite to that above for the CASPEC stars, indicating that
any zero-point correction to our intrinsic-colour equation is not very significant.

\item  And, in addition, as can be seen from panel~(c) of Fig.~3, for the combined
databases of CASPEC, UVES/VLT, and FIES/NOT, the average reddening value is 
$\langle E(b-y) \rangle = -0.001\pm0.002$, which is insignificantly different 
from $E(b$--$y) = 0.00$.  These more recently observed F- and G-type metal-poor
stars help prove that any zero-point correction to our intrinsic-colour equation
must be very small.

\item For the main sequence stars of \astrobj{M67} and \astrobj{NGC~2548}, the estimated average reddenings, 
$+0.020\pm0.004$ for \astrobj{M67} and $+0.060\pm0.011$ for \astrobj{NGC~2548}, are in good concordance 
with the ones found by B05 and B07, and by N87.

\item The intrinsic-colour calibration of our Equation~(1) can be used for
de-reddening $uvby$--$\beta$ photometry for measuring photometric effective
temperatures, metal abundances, absolute magnitudes, distances, stellar
classifications, and ages for dwarf and turn-off stars, in the field and in
clusters of the Galaxy, over a wide range in metallicity.

\end{enumerate}

\section{Acknowledgments}

This work was supported by the CONACyT project 49434-F (M\'exico), and by the Research
Fund of the University of Istanbul, project number: BYP-781/05102005.  We sincerely thank
P.~E.~Nissen for providing his special spectroscopic results, as well as ideas and
references.  This research has been made possible by the use of the SIMBAD database, 
operated at the CDS, Strasbourg, France, and the web site of the General Catalogue of
Photometric Data, Geneva, Switzerland.

\clearpage

\begin{table*}
\tiny
\begin{minipage}{170mm}
\caption{Dependence of Average Residuals upon $m_{1}$, $c_{1}$, and $\beta$ for Equation~(1).
Col.~1: $m_{1}$, $c_{1}$, and $\beta$ intervals, Col.~2: average residuals from Equation~(1),
Col.~3: the number of stars in each interval, Col.~4: standard deviations from Equation~(1).}
{\scriptsize
\begin{tabular}{lclc}
\hline
$m_{1}$ & Ave. residual& N & Std.dev.\\
\hline
{\rm[0.06  0.13]} & $-$0.0013&  24 & 0.0096\\
{\rm(0.13  0.17]} & $-$0.0004&  60 & 0.0099\\
{\rm(0.17  0.21]} & +0.0005  &  73 & 0.0080\\
{\rm(0.21  0.28]} & +0.0003  &  67 & 0.0090\\
{\rm(0.28  0.35]} & +0.0003  &  65 & 0.0088\\
{\rm(0.35  0.45]} & $-$0.0012&  46 & 0.0075\\
{\rm(0.45  0.58]} & +0.0007  &  57 & 0.0085\\
\hline \\
$c_{1}$ & Ave. residual& N & Std.dev.\\
\hline \\
{\rm[0.12  0.25]} &  +0.0000  &  61 & 0.0104\\
{\rm(0.25  0.28]} & $-$0.0006 &  89 & 0.0085\\
{\rm(0.28  0.31]} &   +0.0015 &  84 & 0.0084\\
{\rm(0.31  0.34]} & $-$0.0014 &  77 & 0.0074\\
{\rm(0.34  0.40]} &   +0.0004 &  71 & 0.0087\\
{\rm(0.40  0.51]} & $-$0.0005 &  10 & 0.0108\\
\hline \\
$\beta$ & Ave. residual& N & Std.dev.\\
\hline \\
{\rm[2.488  2.550]} & +0.0006   &  98 & 0.0093\\
{\rm(2.550  2.570]} & $-$0.0015 &  99 & 0.0082\\
{\rm(2.570  2.590]} & +0.0009   & 114 & 0.0081\\
{\rm(2.590  2.610]} & +0.0006   &  58 & 0.0080\\
{\rm(2.610  2.668]} & $-$0.0022 &  23 & 0.0118\\
\hline
\end{tabular} 
}
\end{minipage}
\end{table*}

\begin{table*}
\caption{Mean $\langle E(b$--$y) \rangle$ Values for Two Metal-poor Ranges
and $d < 70$ pc in the Schuster Catalogue, Calculated from Equation~(1).}
\begin{tabular}{lcc}
\hline
$[Fe/H]range$&$\langle E(b$--$y)_{eq.(1)} \rangle$& N\\
\hline
$[Fe/H]<-1.0$&+0.004&21 \\
$[Fe/H]<-1.5$&+0.010& 8 \\
\hline
\end{tabular}  
\end{table*}

\clearpage

\begin{table*}
\caption{Mean $\langle \Delta E(b$--$y) \rangle$ Differences between Equation~(1)
and SN89, or O88, for Metallicity Subsets. Col.~1: $[Fe/H]$ intervals;
Col.~2: the number of stars in each interval;
Col.~3: first part, $\langle \Delta E(b$--$y) \rangle$ differences with respect to SN89,
with (and without) the +0.005 zero-point correction of Nissen (1994); second part, with
respect to O88; Col.~4: standard deviations for the comparisons.}
\begin{tabular}{lccc}
\hline
$[Fe/H]range$& N&$\langle E(b$--$y)_{eq.(1)-SN89}\rangle $&Std.dev.\\
\hline
{\rm[$+$0.42, $+$0.00]} & 19 & $+$0.0057~($+$0.0007) & 0.0074 \\
{\rm($-$0.00, $-$0.25]} & 27 & $+$0.0044~($-$0.0006) & 0.0028 \\
{\rm($-$0.25, $-$0.50]} & 43 & $+$0.0035~($-$0.0015) & 0.0055 \\
{\rm($-$0.50, $-$0.75]} & 46 & $+$0.0062~($+$0.0012) & 0.0033 \\
{\rm($-$0.75, $-$1.00]} & 45 & $+$0.0019~($-$0.0031) & 0.0144 \\
{\rm($-$1.00, $-$1.50]} & 43 & $+$0.0041~($-$0.0009) & 0.0124 \\
{\rm($-$1.50, $-$2.00]} & 21 & $-$0.0030~($-$0.0080) & 0.0184 \\
{\rm($-$2.00, $-$2.72]} & 10 & $+$0.0024~($-$0.0026) & 0.0157 \\
\hline \\
$[Fe/H]$ range & N & $\langle \Delta E(b$--$y)_{eq.(1)- O88} \rangle$ & Std.dev.\\
\hline \\
{\rm[$+0.32$, $-0.50$]} & 342 & $-0.024$ & 0.030 \\
{\rm($-0.50$, $-1.22$]} &  42 & $-0.007$ & 0.007 \\
\hline
\end{tabular}  
\end{table*}

\begin{figure*}
\includegraphics*[width = 8cm, clip=]{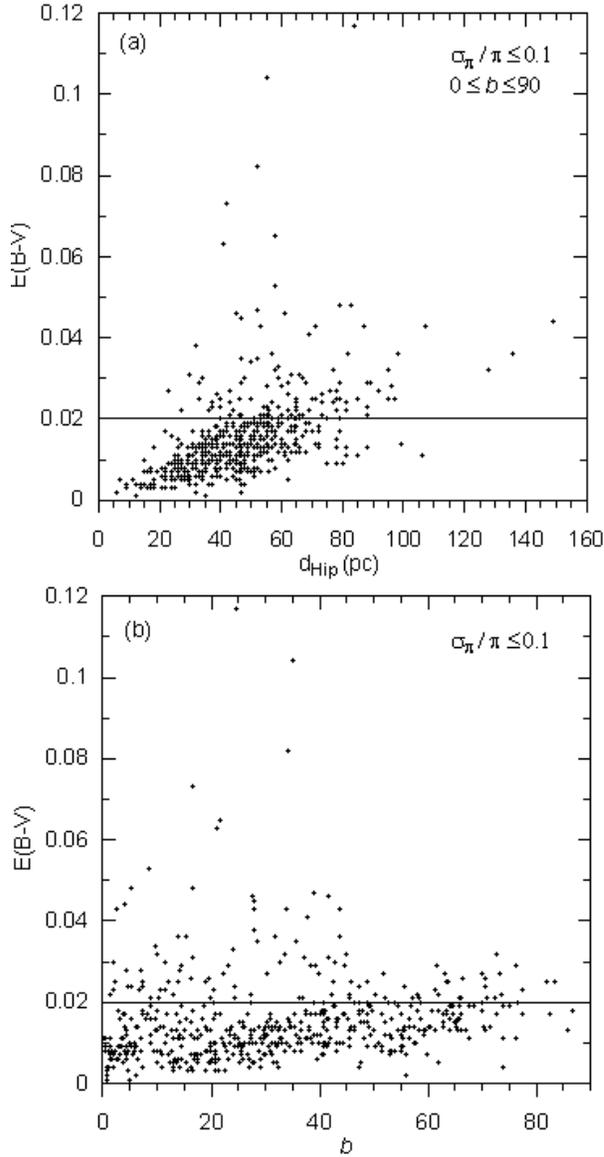}
\caption{For the Schuster catalogue: (a) $E(B$--$V)$ versus $d_{Hip}$; and (b) $E(B$--$V)$
versus Galactic latitude. For this catalogue, the criteria, $E(B$--$V) < 0.02$ plus
$d_{Hip} < 70$ pc have been used to define negligible reddening for the 
intrinsic-colour ($(b$--$y)_{o}$--$\beta$) calibration.  See Section~4 for details.}
\end{figure*}

\begin{figure*}
\includegraphics*[width=7cm, clip=]{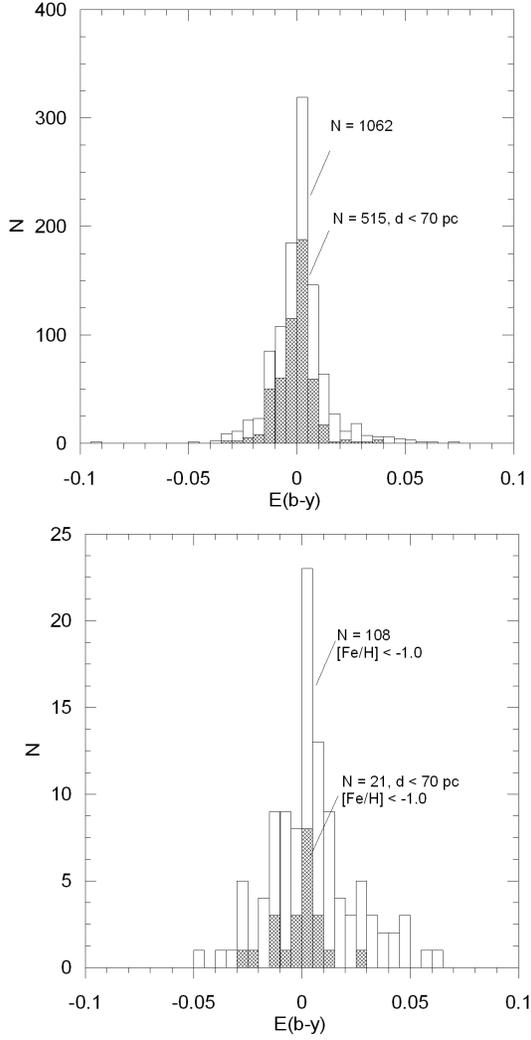}
\caption{$E(b$--$y)$ distributions for stars from the Schuster catalogue are displayed
in panels~(a) and (b) as calculated from Equation~(1); the stars fall within the ranges
of validity:  +$0.290 \leq (b$--$y) \leq +0.606$, $+0.060 \leq m_{1} \leq +0.574$,
$+0.126 \leq c_{1} \leq +0.504$, and $2.488 \leq \beta \leq 2.668$.  Panel (a) includes
1062 stars within the above limits of $(b$--$y)$, $m_{1}$, $c_{1}$, and $\beta$, while
(b) only those stars with $[Fe/H] < -1.0$.  In both panels the distribution is quite
symmetric about $E(b-y)= 0.00$.  The hatched areas meet the restriction, $d_{Hip} < 70$ pc, 
which corresponds to the negligible-reddening, near-solar vicinity.}
\end{figure*}

\begin{figure*}
\includegraphics*[width=7cm,clip=]{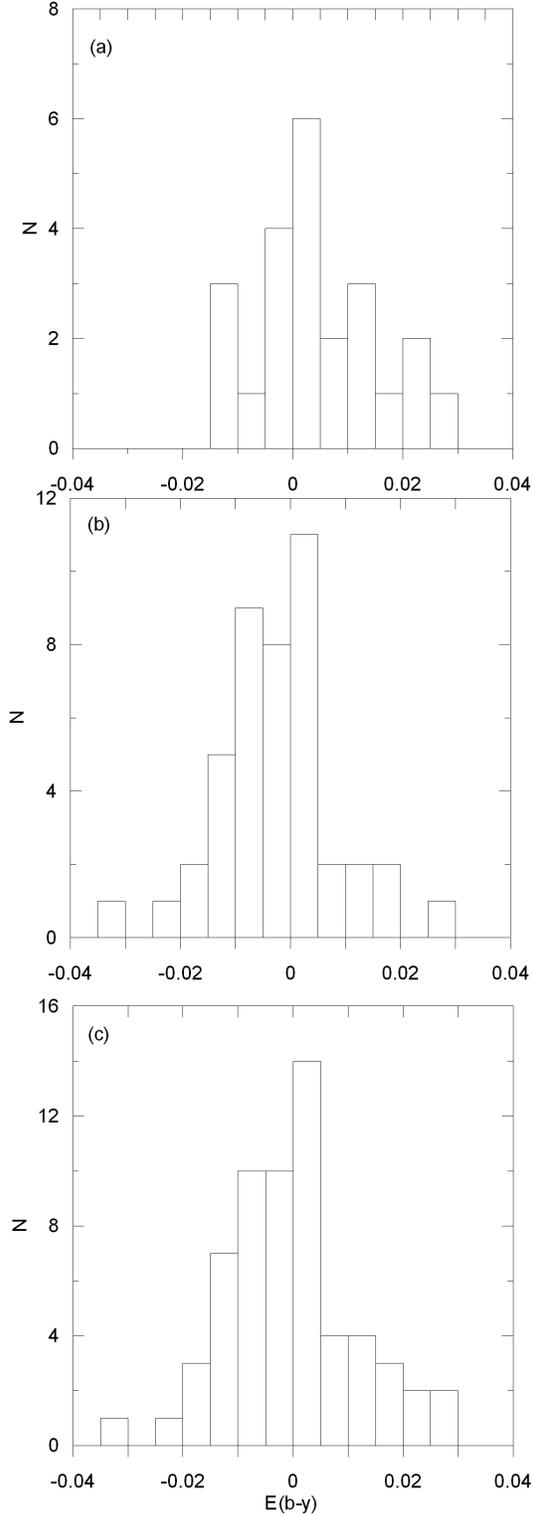}
\caption{$E(b$--$y)$ distributions for:   panel~(a), 23 CASPEC stars;  panel~(b),
44 UVES/VLT + FIES/NOT stars;  and panel~(c), 61 stars, the combined sample; all 
without detectable interstellar Na I lines.  As calculated from our Equation~(1), 
panels~(a), (b), and (c) give the average reddenings of $\langle E(b-y)\rangle  =
+0.004\pm0.002$, $\langle E(b-y) \rangle  = -0.003\pm0.003$, and $\langle E(b-y) \rangle
= -0.001\pm0.002$, respectively.}
\end{figure*}

\end{document}